\documentclass[12pt]{article}
\begin{document}

\title{\bf Energy-Momentum Distribution of the Weyl-Lewis-Papapetrou and
the Levi-Civita Metrics}

\author{M. Sharif \thanks{e-mail: msharif@math.pu.edu.pk} \\
Department of Mathematics, University of the Punjab,\\ Quaid-e-Azam
Campus Lahore-54590, PAKISTAN.}

\date{}

\maketitle

\begin{abstract}
This paper is devoted to compute the energy-momentum densities for
two exact solutions of the Einstein field equations by using the
prescriptions of Einstein, Landau-Lifshitz, Papapetrou and
M\"{o}ller. The spacetimes under consideration are the
Weyl-Lewis-Papapetrou and the Levi-Civita metrics. The Weyl metric
becomes the special case of the Weyl-Lewis-Papapetrou solution. The
Levi-Civita metric provides constant momentum in each prescription
with different energy density. The Weyl-Lewis-Papapetrou metric
yields all the quantities different in each prescription. These
differences support the well-defined proposal developed by
Cooperstock and from the energy-momentum tensor itself.
\end{abstract}


\date{}

\section{Introduction}

The relativistic analogues of the classical principle of the
conservation of energy and momentum can be obtained with the help of
the well-known equation [1]
\begin{eqnarray*}
\frac{\partial}{\partial x^b}(\Im^b_a+t^b_a)=0, \quad (a,b=0,1,2,3),
\end{eqnarray*}
where $\Im^b_a$ is a tensor density of material energy and momentum
and $t^b_a$ is the pseudo-tensor density of gravitational energy and
momentum. The definition of localized energy density is a
longstanding problem [2] in General Relativity (GR). On the basis of
the principle of equivalence, it is usually assumed that the
gravitational energy cannot be localized. The principle of
equivalence is frequently invoked to ensure that the gravitational
field can be made vanish in a sufficiently small region of the
spacetime. Misner at el. [3] showed that the energy can only be
localized in spherical systems. But later on, Cooperstock and
Sarracino [4] proved that if energy is localizable for spherical
systems, then it can be localized in any system. Much attention has
been devoted for this problematic issue.

An energy-momentum complex is the sum of the energy-momentum of
matter and an appropriate pseudo-tensor. Einstein showed that the
energy-momentum pseudo-complex provides satisfactory expression for
the total energy and momentum of a closed system in the form of
three dimensional integral. There are some drawbacks of the Einstein
energy-momentum complex. One of these drawbacks is that it is not
symmetric in its indices. However, Landau-Lifshitz energy-momentum
complex satisfies this requirement. In order to determine the
conserved total four-momentum for gravitation with matter,
Landau-Lifshitz introduced a system of coordinates at some
particular point in spacetime for which all the first derivatives of
the metric tensor vanish. Papapetrou energy-momentum complex is the
least known among the four definitions under discussion and as a
result, it has been re-discovered several times. Although the
Einstein energy-momentum complex provides useful expression for the
total energy-momentum of a closed system. However, from the GR
viewpoint, M\"{o}ller argued that it is unsatisfactory to transform
a system into quasi-Cartesian coordinates. M\"{o}ller tried to find
out an expression of energy-momentum which is independent of the
choice of particular coordinate system.

Einstein was the first to construct a locally conserved
energy-momentum complex [5]. After this attempt, many physicists
including Tolman [6], Landau-Lifshitz [7], Papapetrou [8], Bergmann
[9] and Weinberg [10] introduced different definitions for the
energy-momentum complex. These definitions can only give meaningful
results if the calculations are performed in Cartesian coordinates.
In 1990, Bondi [11] argued that a non-localizable form of energy is
not allowed in GR. After this, the idea of quasi-local energy was
introduced by Penrose and other researchers [12-14]. In this method,
one can use any coordinate system while finding the quasi-local
masses to obtain the energy-momentum of a curved spacetime.
Bergqvist [15] considered seven different definitions of quasi-local
mass and showed that no two of these definitions give the same
result. Chang at el. [16] showed that every energy-momentum complex
can be associated with a particular Hamiltonian boundary term and
hence the energy-momentum complexes may also be considered as
quasi-local.

M\"{o}ller [17,18] proposed an expression which is the best to make
calculations in any coordinate system. He claimed that his
expression would give the same results for the total energy and
momentum as the Einstein's energy-momentum complex for a closed
system. Lessner [19] gave his opinion that M\"{o}ller's definition
is a powerful concept of energy and momentum in GR. However,
M\"{o}ller's prescription was also criticized by some people
[11,20,21]. Komar's complex [21], though not restricted to the use
of Cartesian coordinates, is not applicable to non-static
spacetimes. Thus each of these energy-momentum complex has its own
drawbacks. As a result, these ideas of the energy-momentum complexes
could not lead to some unique definition of energy in GR.

Virbhadra [22] generated interest on this topic by using different
prescriptions to calculate energy-momentum of a spacetime. He found
that different prescriptions could lead to the same result if
appropriate coordinates are used. Aguirregabiria et al. [23] showed
that five different energy-momentum complexes gave the same result
for any Kerr-Schild class (including the Schwarzchild,
Reissner-Nordstr\"{o}m, Kerr and Vaidya metrics). Xulu [24,25]
extended this investigation and found same energy distribution in
the Melvin magnetic and Bianchi type I universe. Chamorro and
Virbhadra [26] and Xulu [27] studied the energy distribution of
charged black holes with a dilaton field.

It was hoped [25] that some particular properties might give a basis
to believe that some pseudo-tensors of energy-momentum density had a
special meaning. However, there exists examples of spacetimes which
do not support this viewpoint. In this regard, Sharif [28,29]
considered the class of gravitational waves and G\"{o}del universe
and used the four definitions of energy-momentum. He concluded that
results obtained are not consistent in different prescriptions.
Recently, Sharif and Fatima [30,31] considered some more examples of
Non-Null Einstein-Maxwell solution, singularity-free cosmological
model and Weyl metrics and applied four different complexes. They
found that the energy-momentum complexes do not provide the same
results for any of these spacetimes. This paper continues the
investigation by considering two more examples.

The paper is organized as follows. In section 2, we shall briefly
mention different prescriptions to evaluate energy-momentum
distribution. Sections 3 and 4 are devoted for the evaluation of
energy-momentum densities for the two particular spacetimes using
the prescriptions of Einstein, Landau-Lifshitz, Papapetrou and
M\"{o}ller. Finally, in the last section, we shall discuss and
summarize all the results obtained.

\section{Energy-Momentum Complexes}

In this section, we shall elaborate four different approaches to
evaluate the energy-momentum density components of different
spacetimes.

\subsection{Einstein Energy-Momentum Complex}

The energy-momentum complex of Einstein [2] is given by
\begin{equation}
\Theta^{b}_{a}=\frac{1}{16\pi}H^{bc}_{a,c},\quad
(a,b,...=0,1,2,3),
\end{equation}
where
\begin{equation}
H^{bc}_a=\frac{g_{ad}}{\sqrt{-g}}[-g(g^{bd}g^{ce}-g^{cd}g^{be})]_{,e}.
\end{equation}
It is to be noted that $H^{bc}_a$ is anti-symmetric in indices $b$
and $c$. $\Theta^{0}_{0}$ is the energy density,
$\Theta^{i}_{0}~(i=1,2,3)$ are the components of momentum density
and $\Theta^{0}_{i}$ are the energy current density components.
Einstein showed that the energy-momentum pseudo-complex
$\Theta^{b}_{a}$ provides satisfactory expression for the total
energy and momentum of closed system in the form of 3-dimensional
integral.

\subsection{Landau-Lifshitz Energy-Momentum Complex}

There were some drawbacks of Einstein energy-momentum complex. One
main drawback was that it was not symmetric in its indices. As a
result, this cannot be used to define conservation laws of angular
momentum. However, Landau-Lifshitz energy-momentum complex is
symmetric and they are able to develop a conserved angular momentum
complex in addition to that of energy-momentum. The energy-momentum
complex of Landau-Lifshitz [7] is given by
\begin{equation}
L^{ab}= \frac{1}{16 \pi}\ell^{acbd}_{,cd},
\end{equation}
where
\begin{equation}
\ell^{acbd}= -g(g^{ab}g^{cd}-g^{ad}g^{cb}).
\end{equation}
$L^{00}$ represents the energy density of the whole system including
gravitation and $L^{oi}$ represent the components of the momentum
density. $\ell^{abcd}$ has symmetries of the Riemann curvature
tensor. It is clear from Eq.(3) that $L^{ab}$ is symmetric with
respect to its indices.

\subsection{Papapetrou Energy-Momentum Complex}

Papapetrou energy-momentum complex is the least known among the four
definitions under discussion, as a result, it has been re-discovered
several times. The expression was found using the generalized
Belinfante method. The symmetric energy-momentum complex of
Papapetrou [8] is given as
\begin{equation}
\Omega^{ab}=\frac{1}{16\pi}N^{abcd}_{,cd},
\end{equation}
where
\begin{equation}
N^{abcd}=\sqrt{-g}(g^{ab}\eta^{cd}-g^{ac}\eta^{bd}
+g^{cd}\eta^{ab}-g^{bd}\eta^{ac}),
\end{equation}
and $\eta^{ab}$ is the Minkowski spacetime. The quantities
$N^{abcd}$ are symmetric in its first two indices $a$ and $b$. The
locally conserved quantities $\Omega^{ab}$ contain contribution from
the matter, non-gravitational and gravitational field. The quantity
$\Omega^{00}$ represents energy density and $\Omega^{0i}$ are the
momentum density components.

\subsection{M\"{o}ller Energy-Momentum Complex}

Although the Einstein energy-momentum complex provides useful
expression for the total energy-momentum of a closed system.
However, from the GR viewpoint, M\"{o}ller [17] argued that it is
unsatisfactory to transform a system into quasi-Cartesian
coordinates. M\"{o}ller tried to find out an expression of
energy-momentum which is independent of the choice of particular
coordinate system. His energy-momentum complex is given by
\begin{equation}
M^{b}_{a}=\frac{1}{8\pi}K^{bc}_{a,c},
\end{equation}
where
\begin{eqnarray}
K^{bc}_{a}=\sqrt{-g}(g_{ad,e}-g_{ae,d})g^{be}g^{cd}.
\end{eqnarray}
Here $K^{bc}_{a}$ is antisymmetric, $M^0_0$ is the energy density,
$M^i_0$ are the momentum density components and $M^0_i$ are the
energy current density components. In the next two sections, we
apply these prescriptions to evaluate energy-momentum distribution
for two particular examples.

\section{Weyl-Lewis-Papapetrou Metric}

The class of stationary axisymmetric solutions of the Einstein field
equations is the appropriate framework for the attempts to include
the gravitational effect of an external source in an exact
analytical manner [32]. At the same time, such spacetimes are of
obvious astrophysical importance, as they describe the exterior of
the body in equilibrium. The complete family of exact solutions
representing accelerating and rotating black holes with possible
electromagnetic charges and a nut parameter is known in terms of a
modified Plebanski-demianski metric. This demonstrates the
singularity and horizon structure of the sources but not that the
complete spacetime describes two causally separated black holes. To
demonstrate this property, the metric was first cast in the
Weyl-Lewis-Papapetrou form. The line element of stationary
axisymmetric spacetime of the Weyl-Lewis-Papapetrou metric is given
by [33]
\begin{equation}
ds^2=e^{2\psi}(dt-\omega
d\phi)^2-e^{2(\gamma-\psi)}(d\rho^2+dz^2)-\rho^2e^{-2\psi}{d\phi}^2,
\end{equation}
where $\omega$ is the angular velocity and $\gamma,~\psi,~ \omega$
are functions of $\rho$ and z only. It is mentioned here this
reduces to the Weyl metric for $\omega=0$. To get meaningful results
in Einstein, Landau-Lifshitz and Papapetrou prescriptions, we
transform this metric into Cartesian coordinates given by
\begin{eqnarray}
ds^2&=&e^{2\psi}{dt}^2+(\omega^2 e^{2\psi}-\rho^2
e^{-2\psi})(\frac{xdy-ydx}{\rho^2})^2
-e^{2(\gamma-\psi)}(\frac{xdx+ydy}{\rho})^2\nonumber\\
&-&2\omega e^{2\psi}(\frac{xdy-ydx}{\rho^2})dt-
e^{2(\gamma-\psi)}dz^2.
\end{eqnarray}

\subsection{Energy-Momentum Densities in Einstein Complex}

The energy-momentum densities of the Weyl-Lewis-Papapetrou metric
can be found by Einstein complex with the components of $H^{bc}_a$
that can be computed by using Eq.(2). When we make use of these
components in Eq.(1), we obtain the following components of energy,
momentum and energy current densities
\begin{eqnarray}
\Theta^0_0&=&\frac{1}{8\pi\rho}[\gamma_\rho(e^{2\gamma}-1)
-\rho\gamma_{\rho\rho}+2\psi_\rho+2\rho\psi_{\rho\rho}-\rho\gamma_{zz}
+2\rho\psi_{zz}\nonumber\\
&+&\frac{\omega^2_\rho e^{4\psi}}{2\rho}+\frac{\omega\omega_\rho
e^{4\psi}}{2\rho}-\frac{\omega\omega_\rho
e^{4\psi}}{2\rho^2}+\frac{2\omega\omega_\rho\psi_\rho
e^{4\psi}}{\rho}],\\
\Theta^0_1&=&\frac{y}{16\pi\rho^2}[(\omega_{\rho\rho}+\omega_{zz})
+2\omega(\gamma_{\rho\rho}+\gamma_{zz})
+2(\omega_\rho\gamma_\rho+\omega_z\gamma_z)\nonumber\\
&-&2\omega\{2(\gamma_{\rho\rho}-2\psi_{\rho\rho})
+(\gamma_{zz}-2\psi_{zz})\}
-2\omega\{(\gamma_\rho-2\psi_\rho)\psi_\rho\nonumber\\
&+&(\gamma_z-2\psi_z)\psi_z\}+\frac{4\omega^2e^{4\psi}}{\rho^2}
(\omega_\rho\psi_\rho+\omega_z\psi_z)\nonumber\\
&+&\frac{\omega^2e^{4\psi}}{\rho^2}(\omega_{\rho\rho}+\omega_{zz})
+\frac{2\omega e^{4\psi}}{\rho}(\omega^2_\rho+\omega^2_z)
-\frac{2\omega_\rho}{\rho}+\frac{2\omega}{\rho^2}\nonumber\\
&+&\frac{\omega_\rho e^{2\gamma}}{\rho} +\frac{2\omega\gamma_\rho
e^{2\gamma}}{\rho} -\frac{2\omega e^{2\gamma}}{\rho^2}
-\frac{2\omega^2\omega_\rho e^{4\psi}}{\rho^3}],\\
\Theta^0_2&=&-\frac{x}{16\pi\rho^2}[(\omega_{\rho\rho}+\omega_{zz})
+2\omega(\gamma_{\rho\rho}+\gamma_{zz})
+2(\omega_\rho\gamma_\rho+\omega_z\gamma_z)\nonumber\\
&-&2\omega\{2(\gamma_{\rho\rho}-2\psi_{\rho\rho})
+(\gamma_{zz}-2\psi_{zz})\}
-2\omega\{(\gamma_\rho-2\psi_\rho)\psi_\rho\nonumber\\
&+&(\gamma_z-2\psi_z)\psi_z\}
+\frac{4\omega^2e^{4\psi}}{\rho^2}(\omega_\rho\psi_\rho+\omega_z\psi_z)
\nonumber\\&+&\frac{\omega^2e^{4\psi}}{\rho^2}(\omega_{\rho\rho}+\omega_{zz})
+\frac{2\omega e^{4\psi}}{\rho}(\omega^2_\rho+\omega^2_z)
-\frac{2\omega_\rho}{\rho}+\frac{2\omega}{\rho^2}\nonumber\\
&+&\frac{\omega_\rho e^{2\gamma}}{\rho} +\frac{2\omega\gamma_\rho
e^{2\gamma}}{\rho} -\frac{2\omega e^{2\gamma}}{\rho^2}
-\frac{2\omega^2\omega_\rho e^{4\psi}}{\rho^3}],\\
\Theta^1_0&=&-\frac{ye^{4\psi}}{16\pi\rho^2}[(\omega_{\rho\rho}+\omega_{zz})
+4(\omega_\rho\psi_\rho+\omega_z\psi_z)],\\
\Theta^2_0&=&\frac{xe^{4\psi}}{16\pi\rho^2}[(\omega_{\rho\rho}+\omega_{zz})
+4(\omega_\rho\psi_\rho+\omega_z\psi_z)],\\
\Theta^0_3&=&0=\Theta^3_0.
\end{eqnarray}

\subsection{Energy-Momentum Densities
in Landau-Lifshitz Complex}

The non-zero components of $\ell^{abcd}$ can be found by using
Eq.(4) and consequently the components of energy and momentum
(energy current) densities in Landau-Lifshitz prescription turn out
to be
\begin{eqnarray}
L^{00}&=&\frac{1}{16\pi\rho^2}[4\rho(\gamma_\rho-\psi_\rho)e^{(\gamma-\psi)}
-4\rho(\gamma_\rho-2\psi_\rho)e^{2(\gamma-2\psi)}\nonumber\\
&-&2\{(\gamma_{\rho \rho}-2\psi_{\rho \rho})
+(\gamma_{zz}-2\psi_{zz})\}e^{2(\gamma-2\psi)}
-4\{(\gamma_\rho-2\psi_\rho)^2\nonumber\\
&+&(\gamma_z-2\psi_z)^2\}e^{(\gamma-2\psi)}
+\frac{2e^{2\gamma}}{\rho^2}\{\omega(\omega_{\rho\rho}+\omega_{zz})
+2(\omega^2_\rho+\omega^2_z)\nonumber\\
&+&4\omega(\omega_\rho\gamma_\rho+\omega_z\gamma_z)
+\omega^2(\gamma_{\rho\rho}+\gamma_{zz})
+2\omega^2(\gamma^2_\rho+\gamma^2_z)\nonumber\\&-&\frac{2\omega\omega_\rho}{\rho}
-\frac{2\omega^2\gamma_\rho}{\rho}+\frac{\omega^2}{\rho^2}\}],\\
L^{10}&=&L^{01}=-\frac{ye^{2\gamma}}{16\pi\rho^2}
[(\omega_{\rho\rho}+\omega_{zz})
+4(\omega_\rho\gamma_\rho+\omega_z\gamma_z)\nonumber\\
&+&2\omega(\gamma_{\rho\rho}+\gamma_{zz})
+4\omega(\gamma^2_\rho+\gamma^2_z)-\frac{\omega_\rho}{\rho}
-\frac{2\omega\gamma_\rho}{\rho}],\\
L^{20}&=&L^{02}=\frac{xe^{2\gamma}}{16\pi\rho^2}
[(\omega_{\rho\rho}+\omega_{zz})
+4(\omega_\rho\gamma_\rho+\omega_z\gamma_z)\nonumber\\
&+&2\omega(\gamma_{\rho\rho}+\gamma_{zz})
+4\omega(\gamma^2_\rho+\gamma^2_z)-\frac{\omega_\rho}{\rho}
-\frac{2\omega\gamma_\rho}{\rho}],\\
L^{30}&=&L^{03}=0.
\end{eqnarray}

\subsection{Energy-Momentum Densities
in Papapetrou Complex}

Here the non-zero components of $N^{abcd}$ are obtained with the
help of Eq.(6). When we make use of these values in Eq.(5), it
yields the following components of energy and momentum (energy
current) densities
\begin{eqnarray}
\Omega^{00}&=&\frac{e^{2\gamma}}{8\pi\rho}[(1-e^{-4\psi})\gamma_\rho
+\{2\psi_\rho-\rho(\gamma_{\rho\rho}-2\psi_{\rho\rho}
+\gamma_{zz}-2\psi_{zz})\}e^{-4\psi}\nonumber\\
&-&2\rho\{(\gamma_\rho-2\psi_\rho)^2+(\gamma_z-2\psi_z)^2\}e^{-4\psi}
+\frac{1}{\rho}(\omega^2_\rho+\omega^2_z)\nonumber\\&+&\frac{2\omega^2}{\rho}
(\gamma^2_\rho+\gamma^2_z)+\frac{\omega}{\rho}
(\omega_{\rho\rho}+\omega_{zz})
+\frac{\omega^2}{\rho}(\gamma_{\rho\rho}+\gamma_{zz})
\nonumber\\&+&\frac{4\omega\omega_\rho}{\rho}
(\omega_\rho\gamma_\rho+\omega_z\gamma_z)
+\frac{2\omega^2}{\rho^3}-\frac{3\omega^2\gamma_\rho}{\rho^2}
-\frac{3\omega\omega_\rho}{\rho^2}],\\
\Omega^{10}&=&\Omega^{01}=-\frac{ye^{2\gamma}}{16\pi\rho^2}
[(\omega_{\rho\rho}+\omega_{zz})
+4(\omega_\rho\gamma_\rho+\omega_z\gamma_z)\nonumber\\
&+&2\omega(\gamma_{\rho\rho}
+\gamma_{zz})+4\omega(\gamma^2_\rho+\gamma^2_z)-\frac{\omega_\rho}{\rho}
-\frac{2\omega\gamma_\rho}{\rho}],\\
\Omega^{20}&=&\Omega^{02}=\frac{xe^{2\gamma}}{16\pi\rho^2}
[(\omega_{\rho\rho}+\omega_{zz})
+4(\omega_\rho\gamma_\rho+\omega_z\gamma_z)\nonumber\\
&+&2\omega(\gamma_{\rho\rho}+\gamma_{zz})
+4\omega(\gamma^2_\rho+\gamma^2_z)-\frac{\omega_\rho}{\rho}
-\frac{2\omega\gamma_\rho}{\rho}],\\
\Omega^{30}&=&\Omega^{03}=0.
\end{eqnarray}

\subsection{Energy-Momentum Densities
in M\"{o}ller Complex}

This prescription does not require the transformation into Cartesian
coordinates. The non-zero components of $K^{bc}_a$ are found from
Eq.(8). Consequently, the components of energy, momentum and energy
current densities become
\begin{eqnarray}
M^0_0&=&\frac{1}{4\pi}(\psi_\rho+\rho\psi_{\rho\rho}+\rho\psi_{zz})
+\frac{e^{4\psi}}{8\pi\rho}[\omega(\omega_{\rho\rho}+\omega_{zz})\nonumber\\
&+&4(\omega_\rho\psi_\rho+\omega_z\psi_z)+(\omega^2_\rho+\omega^2_z)
-\frac{\omega\omega_\rho}{\rho}],\\
M^0_2&=&-\frac{e^{4\psi}}{8\pi\rho}[(\omega^2+\rho^2)
(\omega_{\rho\rho}+\omega_{zz})\nonumber\\
&+&4(\omega^2+\rho^2)(\omega_\rho\psi_\rho+\omega_z\psi_z)
+2\omega(\omega^2_\rho+\omega^2_z)
\nonumber\\&+&4\omega\rho^2(\psi_{\rho\rho}+\psi_{zz})
+4\rho\omega\omega_\rho+\omega_\rho
-\frac{\omega^2\omega_\rho}{\rho}],\\
M^2_0&=&\frac{e^{4\psi}}{8\pi\rho}[\omega_{\rho\rho}+\omega_{zz}
+4(\omega_\rho\psi_\rho+\omega_z\psi_z)
-\frac{\omega_\rho}{\rho}],\\
M^0_1&=&0=M^1_0=M^0_3=M^3_0.
\end{eqnarray}

\section{The Levi-Civita Metric}

The Levi-Civita metric is given by [34]
\begin{eqnarray}
ds^2=\rho^{4s}dt^2-\rho^{4s(2s-1)}(d\rho^2+dz^2)
-\alpha^2\rho^{2(1-2s)}d\phi^2,
\end{eqnarray}
where $\alpha$ is a parameter and $s$ is a charge density parameter.
The following interpretations are somewhat accepted for:\\
$s=0,~\frac{1}{2}$, this becomes locally flat spacetime,\\
$s=0,~\alpha=1$, this reduces to Minkowski spacetime and \\
$s=0,~\alpha\neq1$, we have cosmic string. \\
One of the most interesting metrics of the family of the Weyl
solutions is called $\gamma$-metric, also known as Zipoy-Voorhes
metric [35]. The Levi-Civita metric can be obtained from a family of
the Weyl-metric, i.e., the $\gamma$-metric by taking the limit when
the length of its Newtonian image source tends to infinity. The line
element can be transformed into Cartesian coordinates and is given
by
\begin{eqnarray}
ds^2&=&\rho^{4s}dt^2-\rho^{4s(2s-1)}(\frac{xdx+ydy}{\rho})^2\nonumber\\
&-&\alpha^2\rho^{2(1-2s)}(\frac{xdy-ydx}{\rho^2})^2
-\rho^{4s(2s-1)}dz^2.
\end{eqnarray}

\subsection{Energy-Momentum Densities in Einstein Complex}

Using the components of $H^{bc}_a$, we obtain the components of
energy-momentum
\begin{eqnarray}
\Theta^0_0&=&\frac{s^2\rho^{8s^2-2}}{2\pi\alpha},\\
\Theta^0_i&=&0=\Theta^i_0
\end{eqnarray}
which gives constant momentum.

\subsection{Energy-Momentum Densities
in Landau-Lifshitz Complex}

The non-zero components of $\ell^{abcd}$ lead to the following
components of energy and momentum (energy current) densities in
Landau-Lifshitz complex
\begin{eqnarray}
L^{00}&=&\frac{s}{2\pi}[(2s-1)\rho^{8s^2}-\alpha^2(8s^3-16s^2+9s-1)]\rho^{8s^2-8s-2},\\
L^{i0}&=&0=L^{0i}.
\end{eqnarray}
This also yields momentum constant.

\subsection{Energy-Momentum Densities
in Papapetrou Complex}

Here the components of energy and momentum (energy current)
densities will become
\begin{eqnarray}
\Omega^{00}&=&\frac{s^2\rho^{8s^2-2}}{2\pi\alpha}[1-8\alpha^2(s-1)^2\rho^{-8s}],\\
\Omega^{i0}&=&0=\Omega^{0i}
\end{eqnarray}
which gives constant momentum.

\subsection{Energy-Momentum Densities
in M\"{o}ller Complex}

The energy-momentum densities turn out to be
\begin{eqnarray}
M^b_a&=&0
\end{eqnarray}
giving a constant energy-momentum. We note that all the
prescriptions provide constant momentum for this metric.

\section{Summary and Discussion}

This paper continues the investigation of comparing various
distributions presented in the literature. We have used four
different prescriptions namely Einstein, Landau-Lifshitz, Papapetrou
and M\"{o}ller to calculate energy-momentum densities of two
particular examples. These prescriptions turn out to be a powerful
tool to evaluate energy-momentum for various physical systems.
Although this work does not resolve the longstanding and crucial
problem of the localization of energy in GR, but provides some
information about it through such solutions. The following tables
yield the non-zero components of the energy-momentum densities in
each case. The notation EM has been used for Energy-Momentum.

\newpage
{\bf {\small Table 1(a)}} {\small \textbf{Weyl-Lewis-Papapetrou
Metric: Einstein Complex}}

\vspace{0.2in}
\begin{center}
\begin{tabular}{|c|c|}

\hline {\bf EM Densities}&{\bf Expressions}\\
\hline $\Theta^0_0$ & $\begin{array}{c} \frac{1}{8\pi\rho}\{
\gamma_\rho(e^{2\gamma}-1)
-\rho\gamma_{\rho\rho}+2\psi_\rho+2\rho\psi_{\rho\rho}-\rho\gamma_{zz}\\
+2\rho\psi_{zz}+\frac{\omega^2_\rho
e^{4\psi}}{2\rho}+\frac{\omega\omega_\rho
e^{4\psi}}{2\rho}-\frac{\omega\omega_\rho
e^{4\psi}}{2\rho^2}+\frac{2\omega\omega_\rho\psi_\rho
e^{4\psi}}{\rho}\}\end{array}$\\
\hline $\Theta^0_1$ & $\begin{array}{c}
\frac{y}{16\pi\rho^2}[(\omega_{\rho\rho}+\omega_{zz})
+2\omega(\gamma_{\rho\rho}+\gamma_{zz}) +2(\omega_\rho\gamma_\rho
\\+\omega_z\gamma_z)
-2\omega\{2(\gamma_{\rho\rho}-2\psi_{\rho\rho})
+(\gamma_{zz}-2\psi_{zz})\}\\-2\omega\{(\gamma_\rho-2\psi_\rho)\psi_\rho
+(\gamma_z-2\psi_z)\psi_z\}+\frac{4\omega^2e^{4\psi}}{\rho^2}
(\omega_\rho\psi_\rho\\+\omega_z\psi_z)
+\frac{\omega^2e^{4\psi}}{\rho^2}(\omega_{\rho\rho}+\omega_{zz})
+\frac{2\omega e^{4\psi}}{\rho}(\omega^2_\rho+\omega^2_z)
\\-\frac{2\omega_\rho}{\rho}+\frac{2\omega}{\rho^2}
+\frac{\omega_\rho e^{2\gamma}}{\rho} +\frac{2\omega\gamma_\rho
e^{2\gamma}}{\rho}-\frac{2\omega e^{2\gamma}}{\rho^2}
-\frac{2\omega^2\omega_\rho e^{4\psi}}{\rho^3}]\end{array}$ \\
\hline $\Theta^0_2$ & $\begin{array}{c}
-\frac{x}{16\pi\rho^2}[(\omega_{\rho\rho}+\omega_{zz})
+2\omega(\gamma_{\rho\rho}+\gamma_{zz})
+2(\omega_\rho\gamma_\rho\\+\omega_z\gamma_z)
-2\omega\{2(\gamma_{\rho\rho}-2\psi_{\rho\rho})
+(\gamma_{zz}-2\psi_{zz})\\-2\omega\{(\gamma_\rho-2\psi_\rho)\psi_\rho
+(\gamma_z-2\psi_z)\psi_z\}
+\frac{4\omega^2e^{4\psi}}{\rho^2}(\omega_\rho\psi_\rho\\+\omega_z\psi_z)
+\frac{\omega^2e^{4\psi}}{\rho^2}(\omega_{\rho\rho}+\omega_{zz})
+\frac{2\omega e^{4\psi}}{\rho}(\omega^2_\rho+\omega^2_z)
\\-\frac{2\omega_\rho}{\rho}+\frac{2\omega}{\rho^2}
+\frac{\omega_\rho e^{2\gamma}}{\rho} +\frac{2\omega\gamma_\rho
e^{2\gamma}}{\rho}-\frac{2\omega e^{2\gamma}}{\rho^2}
-\frac{2\omega^2\omega_\rho e^{4\psi}}{\rho^3}]\end{array}$\\
\hline $\Theta^1_0$ & $\begin{array}{c}
-\frac{ye^{4\psi}}{16\pi\rho^2}[(\omega_{\rho\rho}+\omega_{zz})
+4(\omega_\rho\psi_\rho+\omega_z\psi_z)]\end{array}$\\
\hline $\Theta^2_0$ & $\begin{array}{c}
~\frac{xe^{4\psi}}{16\pi\rho^2}[(\omega_{\rho\rho}+\omega_{zz})
+4(\omega_\rho\psi_\rho+\omega_z\psi_z)]\end{array}$\\
\hline
\end{tabular}
\end{center}
\vspace{0.2in}
{\bf {\small Table 1(b)}} {\small
\textbf{Weyl-Lewis-Papapetrou Metric: Landau-Lifshitz Complex}}
\vspace{0.2in}
\begin{center}
\begin{tabular}{|c|c|}

\hline{\bf EM Densities}&{\bf Expressions}\\

\hline $L^{00}$& $\begin{array}{c}
\frac{1}{16\pi\rho^2}[4\rho(\gamma_\rho-\psi_\rho)e^{(\gamma-\psi)}
-4\rho(\gamma_\rho-2\psi_\rho)e^{2(\gamma-2\psi)} \\
-2\{(\gamma_{\rho\rho}-2\psi_{\rho\rho})
+(\gamma_{zz}-2\psi_{zz})\}e^{2(\gamma-2\psi)} \\
-4\{(\gamma_\rho-2\psi_\rho)^2
+(\gamma_z-2\psi_z)^2\}e^{(\gamma-2\psi)}
+\frac{2e^{2\gamma}}{\rho^2}\{\omega(\omega_{\rho\rho}\\+\omega_{zz})
+2(\omega^2_\rho+\omega^2_z)
+4\omega(\omega_\rho\gamma_\rho+\omega_z\gamma_z)
+\omega^2(\gamma_{\rho\rho}\\+\gamma_{zz})
+2\omega^2(\gamma^2_\rho+\gamma^2_z)-\frac{2\omega\omega_\rho}{\rho}
-\frac{2\omega^2\gamma_\rho}{\rho}+\frac{\omega^2}{\rho^2}\}]\end{array}$\\
\hline $L^{10}=L^{01}$ & $\begin{array}{c}
-\frac{ye^{2\gamma}}{16\pi\rho^2} [(\omega_{\rho\rho}+\omega_{zz})
+4(\omega_\rho\gamma_\rho+\omega_z\gamma_z)\\+2\omega(\gamma_{\rho\rho}
+\gamma_{zz})
+4\omega(\gamma^2_\rho+\gamma^2_z)-\frac{\omega_\rho}{\rho}
-\frac{2\omega\gamma_\rho}{\rho}]\end{array}$\\
\hline $L^{20}=L^{02}$ & $\begin{array}{c}
\frac{xe^{2\gamma}}{16\pi\rho^2} [(\omega_{\rho\rho}+\omega_{zz})
+4(\omega_\rho\gamma_\rho+\omega_z\gamma_z) \\
+2\omega(\gamma_{\rho\rho}+\gamma_{zz})
+4\omega(\gamma^2_\rho+\gamma^2_z)-\frac{\omega_\rho}{\rho}
-\frac{2\omega\gamma_\rho}{\rho} ] \end{array}$  \\
\hline
\end{tabular}
\end{center}

\newpage

{\bf {\small Table 1(c)}} {\small \textbf{Weyl-Lewis-Papapetrou
Metric: Papapetrou Complex}}

\vspace{0.2in}
\begin{center}
\begin{tabular}{|c|c|}

\hline{\bf EM Densities}&{\bf Expressions}\\

\hline $\Omega^{00}$& $\begin{array}{c}
\frac{e^{2\gamma}}{8\pi\rho}[(1-e^{-4\psi})\gamma_\rho
+\{2\psi_\rho-\rho(\gamma_{\rho\rho}-2\psi_{\rho\rho}
\\+\gamma_{zz}-2\psi_{zz})\}e^{-4\psi}
-2\rho\{(\gamma_\rho-2\psi_\rho)^2\\+(\gamma_z-2\psi_z)^2\}e^{-4\psi}
+\frac{1}{\rho}(\omega^2_\rho+\omega^2_z)+\frac{2\omega^2}{\rho}
(\gamma^2_\rho+\gamma^2_z)\\+\frac{\omega}{\rho}
(\omega_{\rho\rho}+\omega_{zz})
+\frac{\omega^2}{\rho}(\gamma_{\rho\rho}+\gamma_{zz})
\\+\frac{4\omega\omega_\rho}{\rho}
(\omega_\rho\gamma_\rho+\omega_z\gamma_z)
+\frac{2\omega^2}{\rho^3}-\frac{3\omega^2\gamma_\rho}{\rho^2}
-\frac{3\omega\omega_\rho}{\rho^2}]\end{array}$\\
\hline $\Omega^{10}=\Omega^{01}$& $\begin{array}{c}
-\frac{ye^{2\gamma}}{16\pi\rho^2}[(\omega_{\rho\rho}+\omega_{zz})
+4(\omega_\rho\gamma_\rho+\omega_z\gamma_z)\\+2\omega(\gamma_{\rho\rho}
+\gamma_{zz})
+4\omega(\gamma^2_\rho+\gamma^2_z)-\frac{\omega_\rho}{\rho}
-\frac{2\omega\gamma_\rho}{\rho}]\end{array}$\\
\hline $\Omega^{20}=\Omega^{02}$& $\begin{array}{c}
\frac{xe^{2\gamma}}{16\pi\rho^2} [(\omega_{\rho\rho}+\omega_{zz})
+4(\omega_\rho\gamma_\rho+\omega_z\gamma_z)
\\+2\omega(\gamma_{\rho\rho}+\gamma_{zz})
+4\omega(\gamma^2_\rho+\gamma^2_z)-\frac{\omega_\rho}{\rho}
-\frac{2\omega\gamma_\rho}{\rho}]\end{array}$\\
\hline
\end{tabular}
\end{center}
\vspace{0.2in}
{\bf {\small Table 1(d)}} {\small
\textbf{Weyl-Lewis-Papapetrou Metric: M\"{o}ller Complex}}
\vspace{0.2in}
\begin{center}
\begin{tabular}{|c|c|}

\hline{\bf EM Densities}&{\bf Expressions}\\
\hline $M^0_0$& $\begin{array}{c}
\frac{1}{4\pi}(\psi_\rho+\rho\psi_{\rho\rho}+\rho\psi_{zz})
+\frac{e^{4\psi}}{8\pi\rho}[\omega(\omega_{\rho\rho}+\omega_{zz})
\\+4(\omega_\rho\psi_\rho+\omega_z\psi_z)+(\omega^2_\rho+\omega^2_z)
-\frac{\omega\omega_\rho}{\rho}]\end{array}$\\
\hline $M^0_2$& $\begin{array}{c}
-\frac{e^{4\psi}}{8\pi\rho}[(\omega^2+\rho^2)
(\omega_{\rho\rho}+\omega_{zz})
+4(\omega^2+\rho^2)(\omega_\rho\psi_\rho\\+\omega_z\psi_z)
+2\omega(\omega^2_\rho+\omega^2_z)
+4\omega\rho^2(\psi_{\rho\rho}+\psi_{zz})
\\+4\rho\omega\omega_\rho+\omega_\rho
-\frac{\omega^2\omega_\rho}{\rho}]\end{array}$\\
\hline $M^2_0$&$
\begin{array}{c}
\frac{e^{4\psi}}{8\pi\rho}[\omega_{\rho\rho}+\omega_{zz}
+4(\omega_\rho\psi_\rho+\omega_z\psi_z)
-\frac{\omega_\rho}{\rho}]\end{array}$\\
\hline
\end{tabular}
\end{center}

\vspace{0.2in}

{\bf {\small Table 2}} {\small \textbf{Levi-Civita Metric}}
\vspace{0.2in}
\begin{center}
\begin{tabular}{|c|c|}
\hline{\bf Prescription}&{\bf Energy-Momentum Densities}\\
\hline Einstein & $
\Theta^0_0=\frac{s^2\rho^{8s^2-2}}{2\pi\alpha},\quad \Theta^0_i=0=\Theta^i_0$\\
\hline Landau-Lifshitz & $\begin{array}{c}
L^{00}=\frac{1}{16\pi}[\rho^{16s^2-8s-2}(16s^2-8s)\\
-2\alpha^2(16s^2-16s-1)\rho^{8s^2-8s-2}]\\L^{i0}=0=L^{0i}\end{array}$\\
\hline Papapetrou &$
\begin{array}{c}
\Omega^{00}=\frac{s^2\rho^{8s^2-2}}{2\pi\alpha}[1-8\alpha^2(s-1)^2\rho^{-8s}]\\
\Omega^{i0}=0=\Omega^{0i}\end{array}$\\
\hline M\"{o}ller & $M^b_a=0$\\
\hline
\end{tabular}
\end{center}
From these tables, it is concluded that the energy-momentum
densities turn out to be finite and well-defined in all the
prescriptions for the spacetimes under consideration. In
Weyl-Lewis-Papapetrou metric, the non-vanishing momentum densities
turn out to be the same in Landau-Lifshitz and Papapetrou complexes
while the energy is different in each complex. It is worth
mentioning here that for $\omega=0$ the results reduce to the case
of the Weyl metric as found in the paper [31]. The energy for the
Levi-Civita metric is different while momentum becomes constant in
each prescription. It is worth mentioning that energy-momentum
becomes constant for $s=0$ as expected for Minkowski spacetime.

We would like to remark that the results of energy-momentum
distribution for different spaceimes are not surprising. They
support the fact that different energy-momentum complexes, which are
pseudo-tensors, are not covariant objects. This is in accordance
with the equivalence principle [3] which implies that the
gravitational field cannot be detected at a point. In GR, many
energy-momentum expressions (reference frame dependent
pseudo-tensors) have been proposed. There is no consensus as to
which is the best. However, each expression has a geometrically and
physically clear significance associated with the boundary
conditions. The difference of results supports the well-defined
proposal developed by Cooperstock [36] and verified by many authors
[28-31,37]. It is mentioned here that the results of the
Weyl-Lewis-Papapetrou metric found in teleparallel theory of gravity
do not coincide with the results in GR [38].

Finally, we would like to mention that Virbhadra found [22]
energy-momentum distribution of the Kerr-Newman metric by using
Einstein, Landau-Lifshitz, Tolman and M$\ddot{o}$ller
energy-momentum complexes. He concluded that the four prescriptions
could give the same result for the Kerr-Newman spacetime if
appropriate coordinates are used. As is well-known, Einstein,
Landau-Lifshitz, Tolman's prescriptions can give meaningful results
only if Cartesian coordinates are used but M$\ddot{o}$ller's
prescription does not require any such condition. We have followed
the same coordinate system to obtain the energy-momentum for the two
spacetimes. One can recover the results only if coordinates can be
defined such that the metric can be made compatible. This is not
always possible.

\vspace{0.1cm}
\begin{description}
\item  {\bf Acknowledgment}
\end{description}

We would like to thank Mr. Jamil Amir for useful discussions on the
subject during its write up.

\vspace{0.5cm} {\bf \large References}

\begin{description}
\item{[1]} Tolman, R.C.: Phys. Rev. \textbf{35}(1930)8.

\item{[2]} Komar, A.: Phys. Rev. \textbf{127}(1962)1411; ibid
\textbf{129}(1963)1873;\\ Penrose, R.: \textit{Proc. R. Soc. London}
\textbf{A381}(1982)53; \textit{In Asymptotic Behavior of Mass and
Spacetime Geometry}, ed. Flaherty, J.F. (Springer, Berlin, 1984).

\item{[3]} Misner, C.W., Thorne, K.S. and Wheeler, J.A.:
\textit{Gravitation} (Freeman, New York, 1973).

\item{[4]} Cooperstock, F.I. and Sarracino, R.S.: J. Phys. A:
Math. Gen. \textbf{11}(1978)877.

\item{[5]} Trautman, A.: \textit{Gravitation: An Introduction to Current Research,}
ed. Witten, L. (Wiley, New York, 1962).

\item{[6]} Tolman, R.C.: \textit{Relativity, Thermodynamics and Cosmology}
(Oxford University Press, London, 1934).

\item{[7]} Landau, L.D. and Lifshitz, E.M.: \textit{The Classical Theory of
Fields} (Pergamon, Oxford, 1980).

\item{[8]} Papapetrou, A.: \textit{Proc. R . Irish Acad.} \textbf{A52}(1948)11.

\item{[9]} Bergmann, P. G. and Thompson, R.: Phys. Rev.
\textbf{89}(1953)400.

\item{[10]} Weinberg, S.: \textit{Gravitation and Cosmology} (Wiley, New York, 1972).

\item{[11]} Bondi, H.: \textit{Proc. R. Soc. London}
\textbf{A427}(1990)249.

\item{[12]} Penrose, R.: \textit{Proc. R. Soc. London}
\textbf{A388}(1982)457; GR 10 \textit{Conference} eds. Bertotti, B.
de Felice, F. and Pascolini, A. Padova, \textbf{1}(1983)607.

\item{[13]} Brown, J.D. and York, Jr. J.W.: Phys. Rev. \textbf{D47}(1993)1407.

\item{[14]} Hayward, S.A.: Phys. Rev. \textbf{D497}(1994)831.

\item{[15]} Bergqvist, G.: Class. Quantum Grav. \textbf{9}(1992)1753.

\item{[16]} Chang, C.C., Nester, J.M. and Chen, C.: Phys. Rev.
Lett. \textbf{83}(1999)1897.

\item{[17]} M\"{o}ller C.: Ann. Phys. (NY) \textbf{4}(1958)347.

\item{[18]} M\"{o}ller C.: Ann. Phys. (NY) \textbf{12}(1961)118.

\item{[19]} Lessner, G.: Gen. Rel. Grav. \textbf{28}(1996)527.

\item{[20]} Kovacs, D.: Gen. Rel. Grav. \textbf{17}(1985)927;\\ Novotny
J.: Gen. Rel. Grav. \textbf{19}(1987)1043.

\item{[21]} Komar, A.: Phys. Rev. \textbf{113}(1959)934.

\item{[22]} Virbhadra, K.S.: Phys. Rev. \textbf{D41}(1990)1086; ibid
\textbf{42}(1990)1066; ibid \textbf{42}(1990)2919.

\item{[23]} Aguirregabiria, J.M., Chamorro, A. and Virbhadra, K.
S.: Gen. Rel. Grav. \textbf{28}(1996)1393.

\item{[24]} Xulu, S.S.: Int. J. of Mod. Phys.
\textbf{A15}(2000)1979; Mod. Phys. Lett. \textbf{A15}(2000)1151.

\item{[25]} Xulu, S.S.: Astrophys. Space Sci. \textbf{283}(2003)23.

\item{[26]} Chamorro, A. and Virbhardra, K.S.: Int. J. of Mod. Phys.
\textbf{D5}(1994)251.

\item{[27]} Xulu, S.S.: Int. J. of Mod. Phys. \textbf{D7}(1998)773.

\item{[28]} Sharif, M.: Int. J. of Mod. Phys.
\textbf{A17}(2002)1175.

\item{[29]} Sharif, M.: Int. J. of Mod. Phys. \textbf{A18}(2003)4361;
ibid \textbf{19}(2004)1495.

\item{[30]} Sharif, M. and Fatima, T.: Int. J. of Mod. Phys. \textbf{A20}(2005)4309.

\item{[31]} Sharif, M. and Fatima, T.: Nouvo Cimento \textbf{B120}(2005)533.

\item{[32]} D'Inverno, R.: \textit{Introducing Einstein's
Relativity} (Oxford University Press, 1995).

\item{[33]} Stephani, H., Kramer, D., MacCallum, M.A.H.,
Hoenselaers, C. and Hearlt, E.: {\it Exact Solutions of Einstein's
Field Equations} (Cambridge University Press, 2003).

\item{[34]} Konkowski, A.D., Helliwell, M.T. and Wieland, C.:
\textit{Gravitation and Cosmology: Proc. of the Spanish Relativity
Meeting} 2002, ed. A. Lobo (University of Barcelona Press, 2003)193.

\item{[35]} Herrera, L., Paiva, M.F. and Santos, O.N.:
J. Math. Phys. \textbf{40}(1999)4064.

\item{[36]} Cooperstock, F.I.: Annals Phys. \textbf{282}(2000)215;\\
Found. Phys. \textbf{22}(1992)1011; \textbf{33}(2003)1033.

\item{[37]} Bringly, Thomas: Mod. Phys. Lett. \textbf{A17}(2002)157;\\
Found. Phys. \textbf{22}(1992)1011; \textbf{33}(2003)1033.

\item{[38]} Sharif, M. and Amir, M.J.: Mod. Phys. Lett. \textbf{A22}(2007)425.

\end{description}

\end{document}